\documentclass[twocolumn,twoside,slac_two]{revtex4}
\usepackage{graphicx}
\usepackage{fancyhdr}
\begin{document}

\title{HFAG Charm Mixing Averages}


\author{B.~Petersen}
\affiliation{CERN, CH-1211 Genève 23, Switzerland}

\begin{abstract}
Recently the first evidence for charm mixing has been reported by
several experiments. To provide averages of these mixing results and
other charm results, a new subgroup of the Heavy Flavor Averaging
Group has been formed. We here report on the method and results of
averaging the charm mixing results.
\end{abstract}

\maketitle

\thispagestyle{fancy}


\section{Introduction}

Almost since the discovery of charm mesons, mixing of
$D^0-\overline{D}^0$ mesons have been sought in analogy to the
well-known $K^0-\overline{K}^0$ mixing. Due to very effective GIM
suppression, the expected mixing rate in the charm system is much
smaller than for kaons. Only very recently, the BaBar
\cite{Aubert:2007wf} and Belle \cite{Staric:2007dt} collaborations
have reported the first evidence of charm mixing\footnote{Shortly
after the CHARM2007 workshop additional results with evidence for
charm mixing has been reported by the BaBar and CDF collaborations. In
these proceedings we will summarize the status at the time of the
workshop.}.  These results have renewed the interest from the theory
community as the observed mixing rate could be caused by physics
beyond the standard model or at least provide additional constraints
on new physics.

None of the mixing measurement have a significance above four standard
deviations, but several have similar precision for the mixing
parameters. By combining the measurements we therefore obtain more
precise values for the mixing parameters and exclude the no-mixing
hypothesis with larger confidence.  Combining the different mixing
measurements is not completely straightforward, since not all
measurements are sensitive to the same charm mixing parameters. 

The Heavy Flavor Averaging Group (HFAG) in 2006 created a subgroup
with the responsibility of providing averages of charm physics
measurements. One of the high priority tasks of this group is to
combine the charm mixing measurements into world-average values for
the fundamental mixing parameters.  The first average assuming CP
conservation was shown at FPCP \cite{Schwartz:2007fw}.  Besides those
results, we here report the first results of combining mixing
measurements where we allow for CP violation.

\section{Averaging Method}

Mixing is present in the $D^0-\overline{D}^0$ system if the mass
eigenstates, $|D_1\rangle$ and $|D_2\rangle$, differ from the flavor
eigenstates, $|D^0\rangle$ and $|\overline{D}^0\rangle$. Generally one
can write $|D_{1,2}\rangle=p|D^0\rangle\pm q|\overline{D}^0\rangle$.
The variables of fundamental interest are the mass difference, $\Delta
M=M_1-M_2$ and decay width difference, $\Delta\Gamma=\Gamma_1-\Gamma_2$
between the two mass eigenstates. Traditionally, in charm mixing
one uses the dimensionless variables, $x=\Delta M/\Gamma$ and $y=\Delta
\Gamma/2\Gamma$, where $\Gamma$ is the average decay width. CP
violation in mixing or in the interference between mixing and decay
would manifest itself as $|q/p|\neq 1$ and
$\phi=\arg(q/p)\ne0$, respectively\footnote{The phase $\phi$ is for the moment assumed to be independent of decay mode.}. In addition CP
violation could show up in the decay itself giving rise to
decay mode dependent parameters.

Most measurements do not directly measure $(x,y)$. For instance in
mixing measurements using $D^{0}\rightarrow K^+\pi^-$ decays there is
a unknown strong phase, $\delta_{K\pi}$, so the results obtained are
for $x'=x\cos\delta_{K\pi}+y\sin\delta_{K\pi}$ and
$y'=-x\sin\delta_{K\pi}+y\cos\delta_{K\pi}$. In the averaging
procedure, we first combine measurements of the same
parameters to obtain the more precise observables. Most measurements
are performed using likelihood fits and the combination is therefore
performed by multiplying likelihood functions from each measurement and
finding the new maximum. By combining likelihoods, correlations between
observables and possible non-Gaussian tails are taken into account.
For measurements which are not using likelihoods, we construct a
likelihood using symmetrized, Gaussian uncertainties.  To combine
different types of measurements, the different combined likelihoods
are recalculated as a function of $(x,y,\delta_{K\pi})$ minimizing
over any other variables. $\delta_{K\pi}$ is included since there is
both a direct measurement \cite{Asner:2006md} and by combining the
$D^{0}\rightarrow K^+\pi^-$ measurement with the other measurement of
$x$ and $y$, one can also get a precise measurement of
$\delta_{K\pi}$. When plotting confidence contours for $(x,y)$ we
minimize the likelihood over $\delta_{K\pi}$.

The combining of likelihood functions is currently only done for the
CP conserving case. In principle it can be done also for the CP
violating case by simply having two more variables, $|q/p|$ and
$\phi$, in the final likelihood function. Unfortunately not all
likelihoods are currently available for the measurements which allow
for CP violation. A simple combination is therefore performed by
forming a $\chi^2$ of all measurements expressed in terms of the
fundamental mixing parameters.  The $\chi^2$ assumes Gaussian errors,
but correlations between observables in each individual measurement is
taken account by using the full covariance matrix for each result.

\section{CP Conserving Averages}

The following averages were performed by adding log likelihoods from fits
where CP conservation was assumed.

\subsection{Lifetime Ratio Average}

\begin{figure}[h]
\centering \includegraphics[width=80mm]{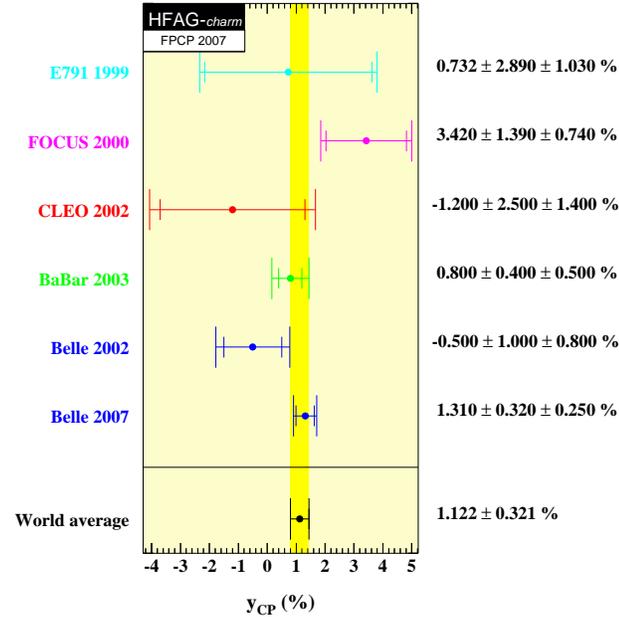}
\caption{Measured $y_{\mathrm{CP}}$ values and the HFAG average.} \label{fig:yCP}
\end{figure}

One can observe charm mixing by finding a difference in the
lifetime measured in decays to CP eigen states such as $D^0\rightarrow
K^+K^-$ and $D^0\rightarrow \pi^+\pi^-$ and the mixed-CP decay
$D^0\rightarrow \pi^+K^-$. We combine six results
\cite{Staric:2007dt,Aitala:1999dt,Link:2000cu,Csorna:2001ww,Aubert:2003pz,Abe:2001ed}
from such analyzes.  All of these measure
$y_{\mathrm{CP}}=\tau_{K\pi}/\tau_{hh}-1$. In the limit of CP
conservation one has $y_{\mathrm{CP}}=y$. The average of the six
measurements is $y_{\mathrm{CP}}=(1.12\pm0.32)\times10^{-2}$. This is
$3.5\sigma$ from the no-mixing hypothesis. As can be seen from
Figure~\ref{fig:yCP}, this average is mainly driven by the recent Belle
measurement.

\subsection{Mixing Rate Average}

\begin{figure}[h]
\centering
\includegraphics[width=80mm]{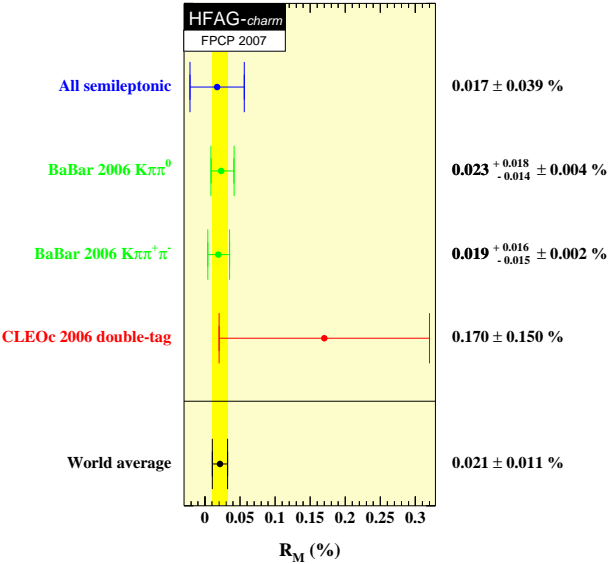}
\caption{The mixing rate from measurements using semileptonic $D^0$ decays are
averaged with results from multi-body hadronic charm decays.} \label{fig:Rm}
\end{figure}

Wrong-signed semileptonic decays provide a clean way of searching for
charm mixing, but the measurements are only sensitive to the
integrated mixing rate $R_M=(x^2+y^2)/2$. Four measurements
\cite{Aitala:1996vz,Cawlfield:2005ze,Abe:2005nq,Aubert:2007aa} are combined and give an
average of $R_M=(1.7\pm3.9)\times10^{-4}$. In addition to the
semileptonic decays, $R_M$ can also measured in the analysis of fully
hadronic decays. The semileptonic result is therefore combined from
two hadronic analyzes \cite{Aubert:2006kt,Aubert:2006rz} and in addition
an analysis of tagged decays at the $\psi(3770)$ \cite{Asner:2006md}. The
combination is illustrated in Figure~\ref{fig:Rm} and gives an average
value of $R_M=(2.1\pm1.1)\time10^{-4}$. In the transformation to a
likelihood in $(x,y)$, we ignore the non-physical region of $R_M<0$.

\subsection{$(x,y)$ Average}

\label{sec:NonKpi}

\begin{figure}[h]
\centering
\includegraphics[width=80mm]{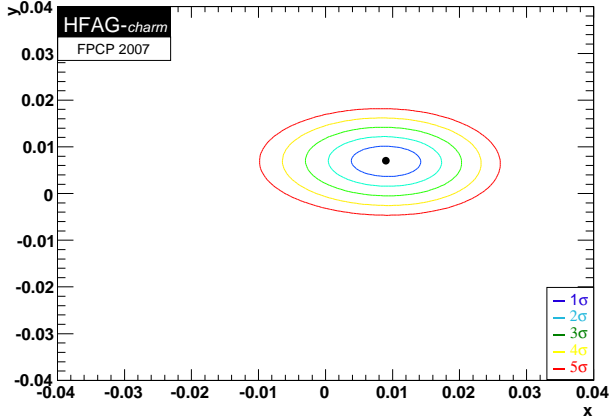}
\caption{Confidence level contours in the mixing parameters $(x,y)$
from the combination of $y_{\mathrm{CP}}$, $R_{M}$
and $(x,y)$ from the time-dependent Dalitz analysis of $D^0\rightarrow
K_S\pi^+\pi^-$.}
\label{fig:xy}
\end{figure}

One can measure $x$ and $y$ directly using a time-dependent Dalitz
plot analysis of $D^0\rightarrow K_S^0\pi^+\pi^-$ decays. Two
measurements \cite{Asner:2005sz,Zhang:2007dt} have been published and
these have been averaged by HFAG and gives $x=(8.1\pm 3.3)\times
10^{-3}$ and $y=(3.1\pm2.8)\times 10^{-3}$. Combining this average
with the averages above for $R_M$ and $y_{\mathrm{CP}}$ using
likelihoods mapped as a function of $(x,y)$ we obtain $x=(9.2\pm
3.4)\times 10^{-3}$ and $y=(7.0\pm2.2)\times 10^{-3}$. Contours of the
combined likelihood function at the levels corresponding to
$1$ to $5\sigma$ confidence levels are shown in
Figure~\ref{fig:xy}. Note that the confidence levels shown correspond
to two-dimensional coverage probabilities of 68.27\%, 95.45\%, etc., and therefore
$2\Delta\ln L=2.30,6.18,$ etc.

\subsection{Averages for $D^0\rightarrow K^+\pi^-$ Decays}

\begin{figure}[h]
\centering
\includegraphics[width=80mm]{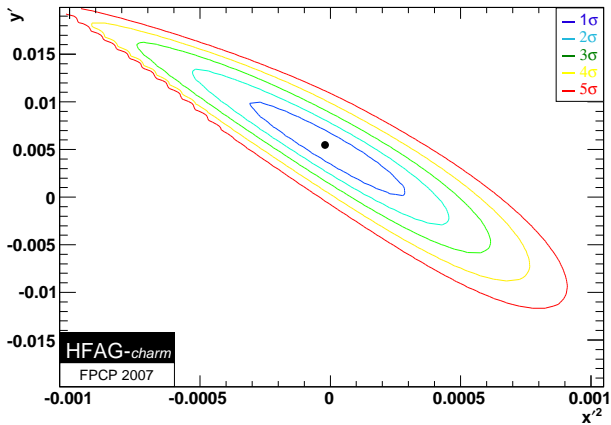}
\caption{Confidence level contours from the combination of BaBar and
Belle measurements using $D^0\rightarrow K^+\pi^-$. The wiggles in the
4 and $5\sigma$ contours for $x'^2<0$ is a binning effect. }
\label{fig:Kpi}
\end{figure}

As mentioned above, one can measure $x'$ and $y'$ using the
doubly-Cabibbo suppressed (DCS) decay $D^0\rightarrow K^+\pi^-$.  The
likelihood functions are available for two measurements
\cite{Aubert:2007wf,Zhang:2006dp} of this type. These are combined and gives
the averages $x'^2=(-0.1\pm2.0)\times10^{-4}$ and
$y'=(5.5^{+2.8}_{-3.7})\times 10^{-3}$. The corresponding likelihood
contours are shown in Figure~\ref{fig:Kpi}. 

\subsection{World Average}

\begin{figure}[h]
\centering
\includegraphics[width=80mm]{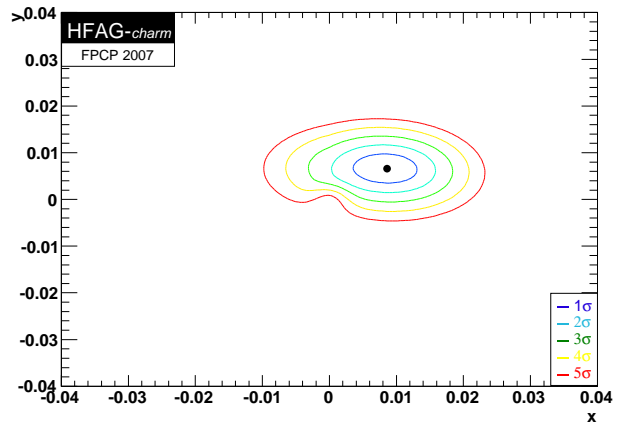}
\caption{Confidence level contours from combining all mixing
measurements under the assumption of CP conservation.} \label{fig:all}
\end{figure}

The combined likelihood for $D^0\rightarrow K^+\pi^-$ decays can be
expressed as a function of $(x,y,\delta_{K\pi})$ ignoring the part
with $x'^2<0$. This likelihood can be combined with the likelihood
from the combination of the other mixing results in
Section~\ref{sec:NonKpi} which do not depend on $\delta_{K\pi}$. An
additional constraint comes from a CLEO-c measurement
\cite{Asner:2006md} of $\cos\delta_{K\pi} = 1.09\pm 0.66$, where a small
dependence on $x$ and $y$ is ignored in the
combination. Figure~\ref{fig:all} shows the likelihood contours in
$(x,y)$ after minimizing over $\delta_{K\pi}$. The region around
the central value is almost unchanged with respect to the result without
the $D^0\rightarrow K^+\pi^-$ decays (Figure~\ref{fig:xy}). This is also
reflected in the over all average for $x$ and $y$ which are
\begin{eqnarray*}
 x&=&(8.7^{+3.0}_{-3.4})\times 10^{-3},\\
 y&=&(6.6\pm2.1)\times 10^{-3}.
\end{eqnarray*}
The $D^0\rightarrow K^+\pi^-$ measurements do not contribute much to
the central value, because of the poorly known phase $\delta_{K\pi}$.
However they do help exclude the no-mixing hypothesis and cause the
dip seen in the contours close to $(x,y)=(0,0)$. At $(x,y)=(0,0)$ we
obtain $2\Delta\ln L=37$ with respect to the minimum. This corresponds to a
significance of the combined mixing signal of $5.7\sigma$.

\begin{figure}[h]
\centering
\includegraphics[width=80mm]{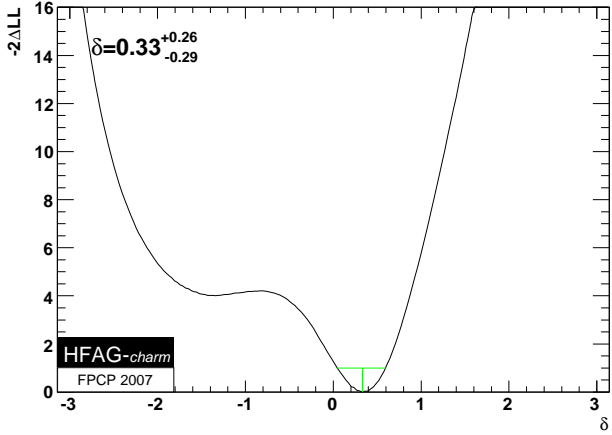}
\caption{Log-likelihood function for the strong phase, $\delta_{K\pi}$,
from combining all mixing measurements. } \label{fig:delta}
\end{figure}

The combination also gives an improved value for $\delta_{K\pi}$. This can
be seen in the projection of the likelihood after minimizing over $x$
and $y$ in Figure~\ref{fig:delta}. The combination gives
$\delta_{K\pi}(=0.33^{+0.26}_{-0.29})\mathrm{rad}$. Without the CLEO-c measurement
of $\delta_{K\pi}$, there would be an equally good second minimum at
$\delta_{K\pi}=-2.17\mathrm{rad}$.

\section{CP Violating Averages}

Measurements of charm mixing can be done without assuming CP
conservation by fitting $D^0$ and $\overline{D}^0$ mesons as separate
samples. Most of the measurements above have done that and we
therefore can combine those to also provide constraints on the CP
violating parameters. When allowing for CP violation, the measured
parameters are related slightly differently to the mixing parameters.
For the lifetime ratio measurements, one has
\begin{eqnarray*}
2y_{\mathrm{CP}}&=&(|q/p|+|p/q|)y\cos\phi-(|q/p|-|p/q|)x\sin\phi,\\
2A_{\Gamma}&=&(|q/p|-|p/q|)y\cos\phi-(|q/p|+|p/q|)x\sin\phi,\\
\end{eqnarray*}
where $A_{\Gamma}$ is the measured relative lifetime difference for
$D^0\rightarrow h^+h^-$ and $\overline{D}^0\rightarrow h^+h^-$.
For $D^0\rightarrow K^+\pi^-$ decays, the $x'$ and $y'$ measured for
$D^0$ and $\overline{D}^0$ are related as follows
\begin{eqnarray*}
x'^{\pm}&=&\left(\frac{1\pm A_M}{1\mp A_M}\right)^{1/4}(x'\cos\phi\pm y'\sin\phi),\\
y'^{\pm}&=&\left(\frac{1\pm A_M}{1\mp A_M}\right)^{1/4}(y'\cos\phi\mp x'\sin\phi),\\
\end{eqnarray*}
where $A_M=\frac{|q/p|^2-|p/q|^2}{|q/p|^2+|p/q|^2}$.  For
$D^0\rightarrow K_S^0\pi^+\pi^-$ decays the measurement directly gives
$x$, $y$, $|q/p|$ and $\phi$, while for the $R_M$ analysis the results
are not separated and therefore just measure $R_M=(x^2+y^2)/2$. The
measurement of $\delta_{K\pi}$ from CLEO-c is not done separately
for $D^0$ and $\overline{D}^0$ mesons and is not included in the
combined result allowing for CP violation.

\begin{figure}[h]
\centering \includegraphics[width=80mm]{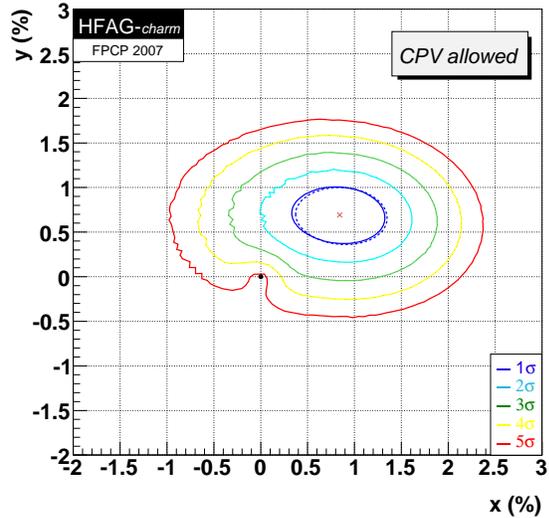}
\caption{Confidence level contours for $(x,y)$ from combining  mixing
measurements with CP violation allowed. The dashed blue curve shows the
$1\sigma$ contour from the CP conserving case for comparison.} \label{fig:xyCPV}
\end{figure}

\begin{figure}[h]
\centering
\includegraphics[width=80mm]{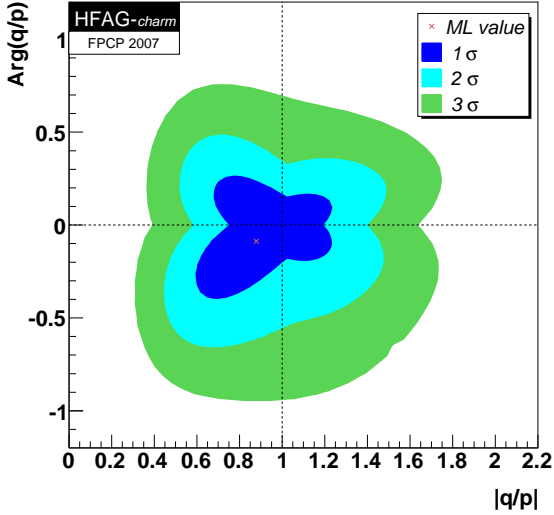}
\caption{Confidence level contours for $(|q/p|,\phi)$ from combining  mixing
measurements with CP violation allowed.} \label{fig:qp}
\end{figure}

In total 22 measurements are combined in a $\chi^2$-fit to extract
seven parameters, the four mixing and CP violation parameters, $x$,
$y$, $|q/p|$ and $\phi$, and three characterizing $D^0\rightarrow K^+\pi^-$,
namely $\delta_{K\pi}$, the DCS rate $R_D$, and the direct decay rate asymmetry
$A_D$. The fit gives $\chi^2=14.4$ and the following mixing parameters
\begin{eqnarray*}
 x&=&(8.4^{+3.2}_{-3.4})\times 10^{-3},\\
 y&=&(6.9\pm2.1)\times 10^{-3},\\
 |q/p|&=&0.88^{+0.23}_{-0.20},\\
 \phi&=&(-0.09^{+0.17}_{-0.19})\,\mathrm{rad}.
\end{eqnarray*}
The mixing parameters are almost unchanged with respect to the CP
conserving average. This is also seen from the confidence
levels shown in Figure~\ref{fig:xyCPV}. One can also draw the
$1$ to $5\sigma$ confidence level contour for $\phi$ versus $|q/p|$ using
$\Delta\chi^2$. This is shown in Figure~\ref{fig:qp}. The no-CP
violation hypothesis is seen to lie well within the $1\sigma$ contour.

The combined results for the $D^0\rightarrow K^+\pi^-$ parameters are
\begin{eqnarray*}
\delta_{K\pi}&=&0.33^{+0.26}_{-0.29}\,\mathrm{rad},\\
R_D&=&(3.35\pm0.11)\times 10^{-3},\\
A_D&=&(-0.8\pm3.1).
\end{eqnarray*}
There is little change in $\delta_{K\pi}$ with respect to the CP
conserving average and no evidence for direct CP violation as $A_D$ is
consistent with zero.

\section{Summary}

Evidence of charm mixing has been reported from several experiments in
the last year. A new subgroup of HFAG has performed an average of
these and other existing charm mixing results. The combined result
has a signal significance in excess of 5 standard deviations and
gives the mixing parameters
\begin{eqnarray*}
 x&=&(8.4^{+3.2}_{-3.4})\times 10^{-3},\\
 y&=&(6.9\pm2.1)\times 10^{-3}.
\end{eqnarray*}
CP violation parameters have also been combined and gives
\begin{eqnarray*}
 |q/p|&=&0.88^{+0.23}_{-0.20},\\
 \phi&=&(-0.09^{+0.17}_{-0.19})\mathrm{rad}.
\end{eqnarray*}
This is fully consistent with no CP violation being present in
charm mixing. HFAG intends to periodically update these averages
as new results become available in order to provide the
most precise mixing parameters to the community.

%\begin{acknowledgments}
%\end{acknowledgments}

\bigskip % extra skip inserted
%\begin{thebibliography}{9}   % Use for  1-9  references

\end{document}